\documentclass{article}
\usepackage{amsmath,amssymb,amsthm,textcomp,color}
\usepackage[all]{xy}

\usepackage[utf8]{inputenc}
\theoremstyle{plain}
\newtheorem{theorem}{Theorem}

\newtheorem{proposition}[theorem]{Proposition}

\theoremstyle{definition}
\newtheorem{definition}[theorem]{Definition}

\voffset = - 1.2cm
\numberwithin{equation}{section}
\numberwithin{theorem}{section}
\input xy
\xyoption{all}

\numberwithin{equation}{section}
\numberwithin{theorem}{section}

\begin{document}

\centerline{{\Large \bf A new application of $k$-symplectic Lie systems}}
\vskip 0.5cm
\centerline{J. de Lucas$^\dagger$, M. Tobolski$^\dagger$ and S. Vilari\~no$^\ddagger$}
\vskip 0.25cm
\centerline{$^\dagger$ Department of Mathematical Methods in Physics, University of Warsaw}
\vskip 0.10cm
\centerline{ul. Pasteura 5, 02-093, Warszawa, Poland}
\vskip 0.25cm
\centerline{$^\ddagger$ Centro Universitario de la Defensa Zaragoza $\&$ IUMA. }
\vskip 0.1cm
\centerline{Academia General Militar. Carretera de Huesca s/n. Zaragoza, 50090, Spain}
 
% \email{$^*$javier.de.lucas@fuw.edu.pl, $^{**}$mariusz.tobolski@student.uw.edu.pl}}

%\author{MARIUSZ TOBOLSKI}
%
%\address{Department of Mathematical Methods in Physics, University of Warsaw,\\ Ul. Pasteura 5,
% 02-093 Warszawa, Poland\\
%\email{mariusz.tobolski@student.uw.edu.pl}}

%\author{SILVIA VILARI\~{N}O}

% \centerline{J.F. Cari\~nena$^\dagger$, J. Clemente-Gallardo$^\dagger$, J.A. Jover-Galtier$^\dagger$ and J. de Lucas$^\ddagger$}
% \vskip 0.25cm
% \centerline{$^\dagger$Faculty of Sciences and IUMA, University of Zaragoza }
% \centerline{c. Pedro Cerbuna 12, 50.009, Zaragoza, Spain. }
% \vskip 0.25cm
% \centerline{$^\ddagger$Department of Mathematical Methods in Physics, University of Warsaw}
% \centerline{ul. Pasteura 5, 02-093, Warszawa, Poland. }

\begin{abstract}

The $k$-symplectic structures appear in the geometric study of the partial differential equations of classical field theories. Meanwhile, we present a new application of the $k$-symplectic structures to investigate a certain type of systems of first-order ordinary differential equations, the $k$-symplectic Lie systems.
In particular, we analyse the properties, e.g. the superposition rules, of a new example of $k$-symplectic Lie system which occurs in the analysis of diffusion equations.

\end{abstract}


\section{Introduction}

The $k$-symplectic structures \cite{Aw-1992} were introduced as a  generalisation of symplectic structures to study geometrically classical field theories. 
Instead of following the standard approach, we introduce a new field of application of $k$-symplectic structures: the Lie systems. A {\it Lie system} is a system of first-order ordinary differential equations whose general solution can be expressed as a function, the \textit{superposition rule}, of a generic finite family of particular solutions and a set of constants. The Lie--Scheffers Theorem \cite{Dissertationes} states that a Lie system amounts to a $t$-dependent vector field taking values in a finite-dimensional Lie algebra of vector fields: a \textit{Vessiot-Guldberg Lie algebra} (VG Lie algebra).

We here focus on Lie systems with VG Lie algebras of Hamiltonian vector fields with respect to a $k$-symplectic structure: the {\it $k$-symplectic Lie systems}. These Lie systems occur in the study of the third-order Kummer-Schwarz equations, Riccati equations, Lie-Lotka-Volterra systems, control theory, etcetera \cite{LV2014}. Moreover, $k$-symplectic structures allow us to devise geometric techniques to obtain superposition rules and other properties of $k$-symplectic Lie systems.

As a main result, we prove that $k$-symplectic Lie systems appear in the study of diffusion equations. Our procedures help in deriving superposition rules and constants of motion for such systems. This shows, as done in \cite{LV2014}, that $k$-symplectic structures can be used to analyse systems of first-order ordinary differential equations. %As a byproduct, we find what it seems a new Casimir element of $\mathfrak{sl}(2,\mathbb{R})\ltimes\mathbb{R}^2$.

\section{$k$-symplectic structures and derived Poisson algebras}

We now recall the notion of $k$-symplectic structures and we relate them to various Poisson algebras (see \cite{LV2014} for details). Mathematical structures are assumed to be real, smooth and globally defined. This leads to omitting technical details and to stress our main results. We hereafter write $\{e^1,\ldots,e^k\}$ for an arbitrary basis of $\mathbb{R}^k$ and $\{ e_1,\ldots,e_k\}$ for its dual one.

\begin{definition}
    Let $N$ be an $n(k+1)$-dimensional manifold and $\omega_1,\ldots,\omega_k$ a set of $k$ closed two-forms on $N$. We say that $(\omega_1,\ldots, \omega_k)$ is a $k$-symplectic structure on $N$ if $\bigcap_{i=1}^k\ker\omega_i(x)=\{0\}$, for all $x\in N$.
\end{definition}
The above definition was introduced by Awane \cite{Aw-1992}. Another generalization of symplectic structures are the polysymplectic structures introduced by G\"{u}nther \cite{Gu-1987}. He defines a {\it $k$-polysymplectic form} on  $N$ as an $\mathbb{R}^k$-valued closed nondegenerated two-form $\Omega=\sum_{i=1}^k\eta_i\otimes e^i$. A $k$-symplectic structure $(\omega_1,\ldots, \omega_k)$ on $N$ gives rise to an associated $k$-polysymplectic form $\Omega=\sum_{i=1}^k\omega_i\otimes e^i$. If $\theta\in (\mathbb{R}^k)^*$, then
$
    \Omega_\theta\equiv \langle \Omega,\theta\rangle =\sum_{i=1}^k\theta(e^i)\omega_i\,
$
is a presymplectic form on $N$. We write ${\rm Adm}\,(\Omega_\theta)$ for the set of admissible functions with respect to the presymplectic manifold $(N,\Omega_\theta)$.
\begin{definition}
Let $(\omega_1,\ldots, \omega_k)$ be a $k$-symplectic structure on $N$ with associated $k$-polysymplectic form $\Omega=\sum_{i=1}^k\omega_i\otimes e^i$. Then,
 \begin{itemize}
 \item A vector field $Y$ on $N$ is $k$-Hamiltonian if it is Hamiltonian relative to  $\omega_1,\ldots, \omega_k$. We write ${\rm Ham}(\Omega)$ for the space of these vector fields.
 \item A function $h=\sum_{\alpha=1}^kh_\alpha\otimes e^\alpha$ is said to be an $\Omega$-Hamiltonian function if there exists a vector field $X_h$ such that $\iota_{X_h}\omega_i=dh_i,\, i=1,\ldots, k$. We denote by $\mathcal{C}^\infty (\Omega)$ the set of $\Omega$-Hamiltonian functions.
 \end{itemize}
\end{definition}

Associated $k$-polysymplectic forms $\Omega$ depends on the chosen bases on $\mathbb{R}^k$. Nevertheless, if $\Omega$ and $\tilde{\Omega}$ are the same up to a change of basis on $\mathbb{R}^k$, all our notions are invariant, up to an eventual change of variables on $\mathbb{R}^k$, e.g. ${\rm Ham}(\Omega)={\rm Ham}(\tilde{\Omega})$.

\begin{proposition}
Let $(\omega_1,\ldots,\omega_k)$ be a $k$-symplectic structure and $\Omega=\sum_{i=1}^k\omega_i\otimes e^i$. 

\begin{itemize}
\item
We have an induced family of Poisson algebras $({\rm Adm}\,(\Omega_\theta),\cdot,\{\cdot,\cdot\}_\theta)$, where $\{\cdot,\cdot\}_\theta$ is the Poisson bracket induced by $\Omega_\theta$ on ${\rm Adm}\,(\Omega_\theta)$ and $\theta$ is any element of $(\mathbb{R}^k)^*$. We call $({\rm Adm}\,(\Omega_\theta),\cdot,\{\cdot,\cdot\}_\theta)$ a \textit{derived Poisson algebra}.
\item The space $\mathcal{C}^\infty(\Omega)$ becomes a Lie algebra with the Lie bracket 
\(
\{h^1,h^2\}_{\Omega}=\sum_{i=1}^k\{h^1_i,h^2_i\}_{\omega_i}\otimes e^i\,,
\)
where $\{\cdot,\cdot\}_{\omega_i}$ is the Poisson bracket induced by $\omega_i$.

\end{itemize}
\end{proposition}

\section{Fundamentals on $k$-symplectic Lie systems}

We now briefly survey the properties of $t$-dependent vector fields, Lie systems and $k$-symplectic Lie systems (see \cite{LV2014} for details).

We denote by $(V,[\cdot,\cdot])$ a Lie algebra given by a vector space $V$ endowed with a Lie bracket $[\cdot\,,\cdot]$.  We define ${\rm
Lie}(\mathcal{B})$ to be  the smallest Lie subalgebra
  of $V$ containing $\mathcal{B}$. We write $V$ instead of  $(V,[\cdot,\cdot])$,  when it is clear what we mean. A $t$-dependent vector field $X$ on $N$ is a $t$-parametric family of vector fields $\{X_t\}_{t\in \mathbb{R}}$.
 There is an evident bijection between $t$-dependent vector fields, e.g. $X=\sum_{i=1}^nX^i(t,x)\partial/\partial x^i$, and systems of the form, namely, ${\rm d}x^i/{\rm d}t=X^i(t,x)$, with $i=1,\ldots,n$. This justifies to use $X$ to represent both a $t$-dependent vector field and its associated system.
 \begin{definition}
        The \textit{minimal Lie algebra} of a $t$-dependent vector field $X$ on $N$ is the smallest real Lie algebra, $V^X$, with respect to the Lie bracket of vector fields containing $\{X_t\}_{t\in \mathbb{R}}$, i.e. $V^X={\rm Lie} (\{X_t\}_{t\in \mathbb{R}},[\cdot,\cdot])\,.$ 
 \end{definition}
The \textit{Lie-Scheffers Theorem} \cite{Dissertationes} asserts  that $X$ admits a superposition rule, i.e. $X$ is a Lie system, if and only if $X$ takes values in a $VG$ Lie algebra, namely $\dim V^X<\infty$. When $V^X$ also consists of Hamiltonian vector fields relative to some structure, numerous results and techniques can be devised to study $X$ \cite{CGLS13,CLS13}. 

\begin{definition}
A \textit{$k$-symplectic Lie system} is a Lie system possessing a VG Lie algebra of $k$-Hamiltonian vector fields relative to a $k$-symplectic structure.
\end{definition}

\begin{definition} Given a vector bundle $\pi:F\rightarrow N$ and a section $\sigma:N\ni n\mapsto \sigma(n)\in F$, we call {\it prolongation} of $\sigma$ to the bundle $\pi^{[m]}:F^{[m]}\equiv F\oplus\ldots\oplus F\ni (f_1,\ldots,f_m)\mapsto (\pi(f_1),\ldots,\pi(f_m))\in N^{m}$ the section $\sigma^{[m]}(f_1,\ldots,f_m)=\sigma(f_1)+\ldots+\sigma(f_m)$.
\end{definition}

 \begin{proposition}Given a $k$-symplectic Lie system $X$ on $N$, let us say $X(t,\xi)=\sum_{\alpha=1}^rb_\alpha(t)X_\alpha(\xi)$, with respect to a $k$-symplectic structure $(\omega_1,\ldots,\omega_k)$, its diagonal prolongation to $N^m$, namely $X(t,\xi_{(1)},\ldots,\xi_{(m)})=\sum_{a=1}^m\sum_{\alpha=1}^rb_\alpha(t)X_\alpha(\xi_{(a)})$, is a $k$-symplectic Lie system relative to $(\omega^{[m]}_1,\ldots,\omega^{[m]}_k)$.
 \end{proposition}
 \begin{proposition}Let $X$ be a $k$-symplectic Lie system on  $N$. For each $\theta\in (\mathbb{R}^k)^*$, the space $\mathcal{I}^X_\theta\!$ of admissible $t$-independent
  constants of motion of $X$   relative to $\Omega_\theta$ is a Poisson algebra with respect to $\{\cdot,\cdot\}_\theta$.
 \end{proposition}

 \begin{theorem}If $X$ is a $k$-symplectic Lie system on $N$ with respect to $(\omega_1,\ldots,\omega_k)$, then there exists a function $h:t\in \mathbb{R}\mapsto h_t\in C^\infty(\Omega)$ such that $\dim {\rm Lie}(\{h_t\}_{t\in\mathbb{R}})<\infty$ and $X_t$ has $\Omega$-Hamiltonian function $h_t$, $\forall t\in\mathbb{R}$. Then, $f\in {\rm Adm}(\Omega_\theta)$ is a constant of motion of $X$ if and only if it Poisson commutes with $ {\rm Lie}(\{\langle h_t,\theta\rangle\}_{t\in\mathbb{R}})$.
 \end{theorem}

\section{Diffusion Riccati system}
Let us consider the following system of differential equations \cite{SSV14} 
\begin{equation*} 
\begin{gathered}
\frac{{\rm d} x}{{\rm d} t}=-a_2(t)+2a_3(t)x+a_1(t)(4x^2+y^4),\,\,
\frac{{\rm d}y}{{\rm d} t}=(a_3(t)+4a_1(t)x)y,\,\,
\frac{{\rm d} z}{{\rm d}t}=a_1(t)y^2,\\
\frac{{\rm d}u}{{\rm d}t}=(a_3(t)+4a_1(t)x)u+a_4(t)-2xa_5(t)+2a_1(t)y^3v,\,\, \frac{{\rm d}v}{{\rm d}t}=-(a_5(t)-2a_1(t)u)y.
\end{gathered}
\end{equation*}
where $a_1(t),\ldots,a_5(t)$ are arbitrary $t$-dependent functions and appearing in the study of diffusion equations. Let us show that the above system is a Lie system related to $X=\sum_{\alpha=1}^5a_\alpha(t)X_\alpha$,
where
$$
\begin{gathered}
X_1=(4x^2+y^4)\frac{\partial}{\partial x}+4xy\frac{\partial}{\partial y}+y^2\frac{\partial}{\partial z}+(4xu+2y^3v)\frac{\partial}{\partial u}+2uy\frac{\partial}{\partial v},\\
X_2=-\frac{\partial}{\partial x},\quad X_3=2x\frac{\partial}{\partial x}+y\frac{\partial}{\partial y}+u\frac{\partial}{\partial u},\quad
X_4=\frac{\partial}{\partial u},\quad X_5=-2x\frac{\partial}{\partial u}-y\frac{\partial}{\partial v}.
\end{gathered}
$$
The vector fields $X_1,\ldots,X_5$ span a Lie algebra isomorphic to $\mathfrak{sl}(2,\mathbb{R})\ltimes \mathbb{R}^2$, indeed $\langle X_1,X_2,X_3\rangle\simeq \mathfrak{sl}(2,\mathbb{R})$ and $\langle X_4,X_5\rangle\simeq \mathbb{R}^2$. It is a long but simple calculation to show that $X_1,\ldots,X_5$ are Hamiltonian relative to the presymplectic structures
$$
\begin{gathered}
\omega_1=\frac{{\rm d}x\wedge {\rm d}y}{y^3},\qquad \omega_\pm=e^{\pm 4z}\left[\frac{{\rm d}x\wedge {\rm d}y}{2y^3}\mp\frac{{\rm d}x\wedge {\rm d}z}{y^2}+\frac{{\rm d}y\wedge {\rm d}z}{y}\right],\\
%:
\omega_2\!=\!\frac{2e^{2z}}{y^2}\left[{\rm d}z\!\wedge\!\left(vy{\rm d}y-2v{\rm d}x+y{\rm d}u-y^2{\rm d}v\right)%\qquad\qquad\qquad \qquad\qquad\qquad \qquad \qquad\qquad\right.
%\left.\qquad \qquad\qquad\qquad \qquad \qquad\qquad
\!+\!\frac{{\rm d}v}2\!\wedge\!(
{y{\rm d}y\!-\!2{\rm d}x})\!+\!\frac{{\rm d}y}{2y}\!\wedge\! \left(4v{\rm d}x\!-\!y{\rm d}u\right)\right]\!.\!
\end{gathered}
$$
Moreover, $\cap_{i=1}^4\ker\omega_i=\{0\}$.  Hence,   $X$ is a $4$-symplectic Lie system with respect to the $4$-symplectic structure  $(\omega_1,\omega_2,\omega_3=\omega_+,\omega_4=\omega_-)$.
We can therefore define the $\Omega$-Hamiltonian function $h_t=\sum_{\alpha=1}^5a_\alpha(t)h^\alpha$, where
$
h^j=\sum_{i=1}^4h^j_i\otimes e^i\,,
$
with $j= 1,2,3,4,5$ and
{\small
$$\begin{array}{|c|c|c|c|}\hline
h^1_1\!\!=\!-\frac{2x^2}{y^2}+\frac{y^2}{2} & h^1_2\!\!=\!\frac{2e^{2z}}{y^2}\left\{4x^2v+uy^3-2vxy^2-2uyx\right\} & h^1_3\!\!=\! e^{4z}\left( x-\frac{x^2}{y^2}-\frac{y^2}{4}\right) & h^1_4\!\!=\! e^{-4z}\left(- x-\frac{x^2}{y^2}-\frac{y^2}{4}\right)\\\hline
h^2_1\!\!=\!\frac{1}{2y^2} & h^2_2\!\!=\!\frac{-2e^{2z}v}{y^2} & h^2_3\!\!=\! \frac{e^{4z}}{4y^2} & h^2_4\!\!=\! \frac{e^{-4z}}{4y^2} \\\hline
h^3_1\!\!=\! \frac{-x}{y^2} & h^3_2\!\!=\!\left(\frac{4vx}{y^2}-\frac{u}{y}-v\right)e^{2z}& h^3_3\!\!=\! \frac{-xe^{4z}}{2y^2}+\frac 14e^{4z} & h^3_4\!\!=\! \frac{-xe^{-4z}}{2y^2}-\frac 14e^{-4z}\\\hline
h^4_1\!\!=\!0& h^4_2\!\!=\!\frac{-e^{2z}}{y} &  h^4_3\!\!=\!0 & h^4_4\!\!=\!0 \\\hline
h^5_1\!\!=\!0& h^5_2\!\!=\!\frac{2xe^{2z}}{y}-ye^{2z} & h^5_3\!\!=\!0 & h^5_4\!\!=\!0 \\\hline
\end{array}
$$}
We have the following non-vanishing commuting relations
$$
\begin{array}{lllllllll}
&\{h^1,h^2\}_\Omega&=-4h^3,\,\,\, &\,\,\,\,\{h^1,h^3\}_\Omega\!\!&=2h_1,\,\,\,  &\{h^1,h^4\}_\Omega&=-2h^5,\,\,\, &\{h^2,h^3\}_\Omega&=-2h^2,\\
& &\{h^2,h^5\}_\Omega&=-2h^4,\quad
&\{h^3,h^4\}_\Omega&=h^4,\quad &\{h^3,h^5\}_\Omega&=-h^5.\quad\\
\end{array}
$$
We have $\langle h^1,\ldots,h^5\rangle\simeq\mathfrak{sl}(2,\mathbb{R})\ltimes\mathbb{R}^2$. If $\theta\in W\equiv\langle e_1,e_3,e_4\rangle$, then $\langle h^1_\theta,\ldots,h_\theta^5\rangle$ is isomorphic to a Lie subalgebra of $\mathfrak{sl}(2,\mathbb{R})$. The above functions help in obtaining superposition rules and constants of motion for $X$. Indeed, we derived that $\mathcal{C}_1^\theta=h_{\theta}^1h_{\theta}^3+(h_{\theta}^2)^2$ for $\theta\in W$ and $\mathcal{C}_2^\theta=(h_{\theta}^4)^2h^1_{\theta}+h^5_{\theta}(2h^4_{\theta}h^3_{\theta}-h^5_{\theta}h^2_{\theta})$ for any $\theta$ are Casimirs of $\mathfrak{sl}(2,\mathbb{R})$ and $\mathfrak{sl}(2,\mathbb{R})\ltimes \mathbb{R}^2$ respectively. From this, we obtain  that
$
\{\mathcal{C}_1^\theta,h_{\theta}^i\}_{\theta}=0,\,\, \theta\in W,\,\,
 \{\mathcal{C}_2^\theta,h_{\theta}^i\}_{\theta}=0,\,\,\theta\in(\mathbb{R}^5)^*,
$
for $i=1,\ldots,5$.
Hence, $\mathcal{C}^\theta_1$ for $\theta\in W$ and $\mathcal{C}_2^\theta$ for any $\theta\in(\mathbb{R}^5)^*$ are $t$-independent constants of motion for $X$. %It is worth noting that $\mathcal{C}_1^\theta$ is a Casimir of .

The diagonal prolongation, $X^{[2]}$, of $X$ to $(\mathbb{R}^5)^2$ is a $k$-symplectic Lie system relative to the $4$-symplectic structure $(\omega^{[2]}_1,\ldots,\omega^{[2]}_4)$. We have that $X^{[2]}_1,\ldots,X^{[2]}_5$ span the same Lie algebra as $X_1,\ldots,X_5$. If we write $\xi_i=(x_i,y_i,z_i,u_i,v_i)\in\mathbb{R}^5$, then $X^{{[2]}}_1,\ldots,X^{[2]}_5$ admit $\Omega$-Hamiltonian functions $h_i^{[2]}(\xi_1,\xi_2)=h_i(\xi_1)+h_i(\xi_2)$ for $i=1,2$. Then, $X^{[2]}$ has $t$-independent constants of motion
$$
\begin{gathered}
\mathcal{C}^{e_1}_1=\frac{-4(x_1-x_2)^2+(y_1^2+y_2^2)^2}{4y_1^2y_2^2},\qquad \mathcal{C}_\pm=\frac{e^{\pm4(z_1+z_2)}[2(x_1-x_2)\mp (y_1^2-y_2^2)]^2}{-16y_1^2y_2^2},\\
\mathcal{C}^{e_1+e_3}_1=\frac{(e^{4z_1}+2)(2(x_1-x_2)+y_2^2)+(2-e^{4z_1})y_1^2}{-16y_1^2y_2^2((e^{4z_2}+2)(2(x_1-x_2)-y_1^2)+(e^{4z_2}-2)y_2^2)^{-1}}
\end{gathered}
$$
and
\begin{multline*}
\mathcal{C}^{e_2}_2 =%\noalign{\medskip}
  \frac{e^{2z_2}[-y_1(u_1-u_2+v_2y_2)+v_1(2(x_1-x_2)+y_2^2)]}{e^{-2(z_1+z_2)}[2(x_1-x_2)-y_1^2+y_2^2]^{-1}y_1^2y_2^2}+\\
\frac{e^{2z_1}[v_2(2(x_1-x_2)-y_1^2)+(-u_1+u_2+v_1y_1)y_2]}{e^{-2(z_1+z_2)}[2(x_1-x_2)-y_1^2+y_2^2]^{-1}y_1^2y_2^2}.
 \end{multline*}
To obtain a superposition rule for $X$ we need as many functionally independent and $t$-independent constants of motion for $X^{[2]}$ as the dimension of $\mathbb{R}^5$, let us say $I_1,\ldots,I_5$ \cite{Dissertationes}. Moreover, we have to require $\partial(I_1,\ldots,I_5)/\partial (x_1,\ldots, v_1)\neq 0$. Our approach provides four of them, e.g. $\mathcal{C}^{e_1}_1,\mathcal{C}_+, \mathcal{C}^{e_1+e_3}_1,\mathcal{C}_2^{e_2}$. An additional constant of motion has to be added to provide the superposition rule for $X$. In any case, it can be proved that $\mathcal{C}^{e_1}_1,\mathcal{C}^{e_3}_1=\mathcal{C}_+, \mathcal{C}^{e_1+e_3}_1$ provide a superposition rule for the projection of $X$ onto $\mathbb{R}^3$.

\section{Conclusion and Outlook}

%We have devised a new application of $k$-symplectic Lie systems: diffusion equations.

By means of $k$-symplectic Lie systems and structures, we analysed superposition rules for a system of first-order ordinary differential equations occurring in the study of diffusion equations.  In the future, we aim to look for new applications of our methods and to expand our techniques to study Lie systems associated with multisymplectic and poly-Dirac structures.

\section*{Acknowledgments}
Research of J. de Lucas is financed by the research project MTM2010-12116-E (Ministerio de Ciencia e Innovaci\'on) and the Polish National Science Centre grant HARMONIA (DEC-2012/04/M/ST1/00523). S. Vilari\~{n}o is partially financed by research projects MTM2011-15725-E and MTM2011-2585 (Ministerio de Ciencia e Innovaci\'{o}n) and E24/1 (Gobierno de Arag\'on).

\end{document}